# CO-LOCATED DIFFUSE APPROXIMATION METHOD FOR TWO DIMENSIONAL INCOMPRESSIBLE CHANNEL FLOWS

C. PRAX and H. SADAT
Laboratoire d'Etudes Thermiques, URA CNRS 1403
40, Avenue du Recteur Pineau 86022 Poitiers Cedex, France

## Introduction

The main contribution of this paper is the formulation of a diffuse approximation method(DAM), for two-dimensional channel flows. The proposed method is based on the vorticity-streamfunction formulation. The DAM which estimates derivates of a scalar field has the remarkable advantage to work on discretization points (thus avoiding mesh generation). It has been shown that the DAM is much better than the finite element method for the computation of gradients [1-2]. In a previous paper [3], we have shown that it can be used to solve laminar natural convection problems. In this work, we discuss the applicability of this method to channel flows with a particular emphasis on the form of the weighting function.

## The Diffuse Approximation

Consider a scalar field $\varphi(x,y)$, in a two dimensionnal domain, and a set of n points $M_i(x_i, y_i)$ in the vicinity of a chosen point $M(x,y)$. The diffuse approximation provides estimates of $\varphi$ and its derivatives at M from the nodal values $\varphi_i$. The starting point is to estimate the Taylor expansion of $\varphi$ at M by a weighted least squares method which uses only the values of $\varphi$ at the nearest points $M_i$. By troncating the series at order k, one obtains the corresponding estimates of the derivatives at the same order.

Let us then estimate the second-order Taylor expansion of $\varphi_i$ at M as:

$$\varphi_i^*(x_i, y_i) = <p(M_i,M)> . <\alpha_M>^T \qquad (1)$$

where $<p(M_i,M)>$ is the line vector of polynomial basis and $<\alpha_M>^T$ is the transposed vector of the approximation defined as :

$$<p(M_i,M)> = <1, (x_i-x), (y_i-y), (x_i-x)^2, (x_i-x).(y_i-y), (y_i-y)^2> \qquad (2)$$

$$<\alpha_M>^T = <\alpha_0, \alpha_1, \alpha_2, \alpha_3, \alpha_4, \alpha_5>^T \qquad (3)$$

The variables $\alpha_M$ are determined by minimizing the quadratic expression:

$$I(\alpha_M) = \sum_{i=1}^{n} \{\omega 1(M,M_i) \cdot [\varphi_i - <p(M_i,M)> \cdot <\alpha_M>^T]^2\} \qquad (4)$$

where $\omega 1(M,M_i)$ is a continuous weighting function, having its maximum value at M and decreasing rapidly to zero. Thus only the nearest points to M are involved in (4). By writing the six conditions:

$$\frac{\partial I(\alpha_M)}{\partial \alpha_j} = 0 \quad , j=0,5 \qquad (5)$$

we obtain the (6x6) linear system:

$$[A^M] \cdot <\alpha_M>^T = <B^M>^T \qquad (6)$$

where

$$[A^M] = \sum_{i=1}^{n} \omega 1(M,M_i) \cdot <p(M_i,M)>^T <p(M_i,M)> \qquad (7)$$

$$<B^M>^T = \sum_{i=1}^{n} \omega 1(M,M_i) \cdot <p(M_i,M)>^T \cdot \varphi_i \qquad (8)$$

Once the system (6) has been solved, one finally obtains the desired estimates of the derivatives at M:

$$\varphi(x,y) = \alpha_0; \quad \frac{\partial \varphi}{\partial x} = \alpha_1; \quad \frac{\partial \varphi}{\partial y} = \alpha_2;$$
$$\frac{\partial^2 \varphi}{\partial x^2} = \alpha_3; \quad \frac{\partial^2 \varphi}{\partial y \partial x} = \alpha_4; \quad \frac{\partial^2 \varphi}{\partial y^2} = \alpha_5; \qquad (9)$$

The weighting function can be chosen in many ways. Its radius must be large enough to overlap at least a number of nodes equal to the number of terms $\alpha_i$. However, the situation where the selected nodes are aligned must be avoided in order to get a non singular $[A^M]$ matrix.

Implementation of the Diffuse Approximation

We consider the Navier-Stokes equations in the vorticity-streamfunction formulation :

$$\Delta \Psi + \omega = 0 \qquad (10)$$

$$\frac{\partial u\omega}{\partial x} + \frac{\partial v\omega}{\partial y} = \nu \Delta \omega \qquad (11)$$

Where $\nu$ is the kinematic viscosity. Our method for solving the equations (10) and (11) by using the diffuse approximation has a remote similarity with the finite difference method. The partial derivatives at a given point are expressed as functions of the neighbouring nodal values of $\Psi$ or $\omega$ (by inverting the matrix $[A^M]$):

$$\left\langle \varphi, \frac{\partial \varphi}{\partial x}, \frac{\partial \varphi}{\partial y}, \frac{\partial^2 \varphi}{\partial x^2}, \frac{\partial^2 \varphi}{\partial y \partial x}, \frac{\partial^2 \varphi}{\partial y^2} \right\rangle^T = \left[ A^M \right]^{-1} \cdot \left\{ \sum_{i=1}^{n} \omega_l(M,M_i) \cdot \left\langle p(M,M_i) \right\rangle^T \cdot \varphi_i \right\} \qquad (12)$$

By using the relations (12), the governing equations (10-11) are now replaced (at every node M) by algebraic expressions in terms of the neighbouring nodal values $\Psi_i$ or $\omega_i$. Two systems are then obtained and solved iteratively after the introduction of the boundary conditions. In this work a relaxation factor of 0.2 is used for each variable and the convergence criteria include the relative changes between consecutive iterations:

$$\left| \frac{\Psi_{new} - \Psi_{old}}{\Psi_{new}} \right|_{max} \leq 10^{-3} \quad ; \quad \left| \frac{\omega_{new} - \omega_{old}}{\omega_{new}} \right|_{max} \leq 10^{-3} \qquad (13)$$

The boundary condition for the streamfunction, at the outflow of the channel, is:

$$\frac{\partial \Psi}{\partial x} = 0$$

The vorticity values at the boundary are calculated in terms of the neighbouring streamfunction values by using the method of Kettleborough et al. [4].

## Applications:

### Flow between Parallel Plates

To evaluate the accuracy of the method, the developing laminar flow between two parallel plates was computed. An uniform velocity is imposed at the inlet section, and a parabolic profile is expected to form at about 0.04 Re [5], where Re is the Reynolds number referred to the width of the channel. Calculations were made for a channel with length-to-width ratio L/D=10. We first considered a 101*11 grid and the following gaussian window:

$$\omega_1(M,M_i) = \text{Exp}\left(-21.\left(\frac{r}{\sigma}\right)^2\right)$$
$$\omega_1(M,M_i) = 0 \quad \text{if } r^2 > \sigma^2 \tag{14}$$

where $r^2 = (x_i-x)^2 + (y_i-y)^2$

We then used a 51*21 grid where the node density ratio is equal to four ($\Delta x = 4\Delta y$). In this case, the previous window selects more points in the transverse direction than in the streamwise direction and the method fails to converge. We have consequently modified the gaussian window by setting:

$$r^2 = \frac{(x_i-x)^2}{4^2} + (y_i-y)^2$$

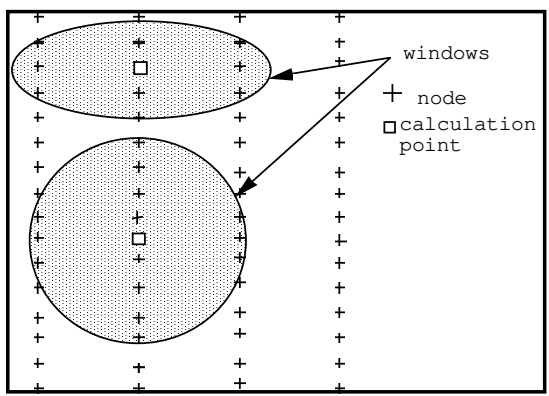
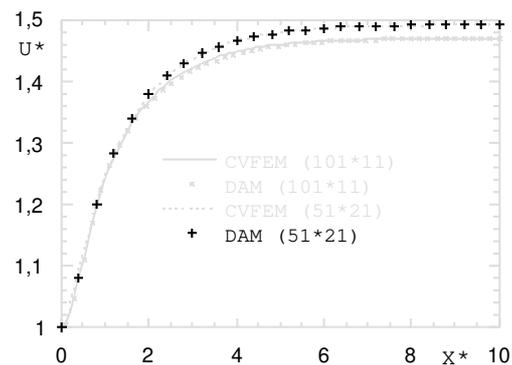

FIG. 1            FIG. 2

Weighting functions      Centerline velocity

These two windows are schematically represented on Fig.1. The calculated centerline velocity as a function of the distance from the inlet for a Reynolds number of 100 is shown on Fig.2. The second mesh gives better results as expected. The numerical value in the fully developed region agrees well with the analytical value of 1.5. In order to compare the DAM with a classical method, we have also reported on FIG.2, the results obtained by using a control volume finite element method (CVFEM)[6]. It appears that the results obtained by the DAM are comparable to those obtained by the well established CVFEM.

Laminar Flow over a Backward Facing Step

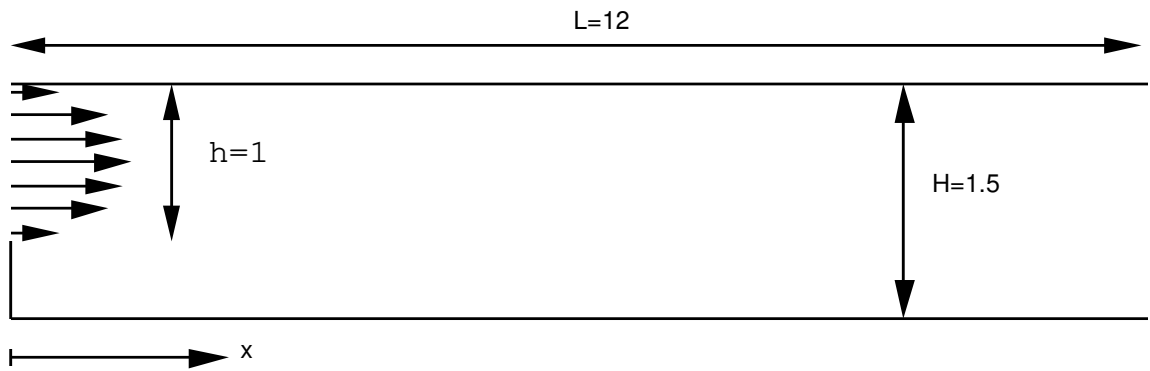

FIG. 3
Backward facing step

Of concern here is the laminar flow over a facing step (FIG.3). A fully developed parabolic laminar flow is prescribed at the inflow section. The governing equations are nondimensionalized by defining

$$X = \frac{x}{H-h} \quad ; \quad Y = \frac{y}{H-h} \quad ; \quad U = \frac{u}{U_{max}} \quad ; \quad V = \frac{v}{U_{max}} \quad ; \quad Re = \frac{U_{max} \cdot (H-h)}{\nu}$$

where $U_{max}$ is the maximum velocity at the inflow section. Calculations were performed until X=24 for Re=50 and Re=150 on two different grids (62*11 and 86*16) by using a gaussian window. In Table 1, the present calculated reatachment length is compared with the results obtained in [7] with a (62*50) mesh. The results obtained by using the CVFEM on a (62*11) grid are also reported on Table 1 for comparison. In FIG.4, the axial velocity profiles are presented. We can see a rather good agreement with the reference and with the CVFEM results.

TABLE 1.

| | Reynolds number | |
|---|---|---|
| | 50 | 150 |
| SOU (62*50) | 2.95 | 6.52 |
| DAM (62*11) | 2.86 | 6.20 |
| CVFEM(62*11) | 2.89 | 6.17 |
| DAM (86*16) | 2.92 | 6.66 |

Reattachement    lengh, $x_r/(H-h)$

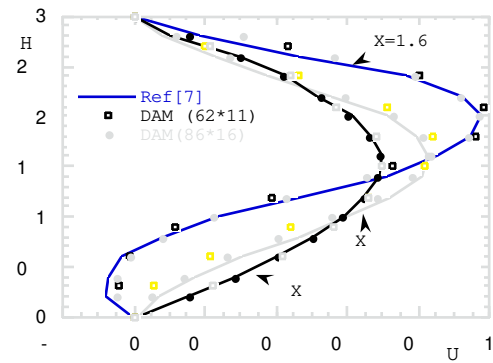

FIG. 4
Axial velocity profiles

## Conclusion

The diffuse approximation method has been applied to fluid flow in channels and compared with a control volume finite element method. Its accuracy has been shown on two test cases. Finally, we have shown that it is better to use an elongated weighting function when the node density is greater in one direction.